# The false myth of the surge in Italian self-citations, and the positive effect of bibliometric evaluations on the increase of the impact of Italian research

Pietro D'Antuono[1,2,*], Michele Ciavarella[1]

[1]Department of Mechanics, Mathematics & Management, Polytechnic University of Bari, Bari BA, Italy

[2]Leonardo Helicopters Division Cascina Costa, Fatigue Office, Samarate, VA, Italy

*Corresponding author
E-mail: mciava @poliba.it (M. Ciavarella)

# Abstract

It has recently been claimed by Baccini and coauthors that due to ANVUR's bibliometric evaluations of individuals, departments, and universities, in Italy there has been a surge in self-citations in the last ten years, thus increasing the "inwardness" of Italian research more than has happened abroad. We have studied the database of Ioannidis et al. published on 12 August 2019 of the one hundred thousand most "highly cited" scientists, including about two thousand Italians, and we found that the problem of self-citations in relation to this scientific elite is not significant in Italy, while perhaps observing a small deviation in the low scores in the rankings. The effect indicated by Baccini et al. consequently, does not seem worrying for the scientific elite (we quantified it in 2% of the total of scientists of the "best" one hundred thousand), and is probably largely concentrated in the further less cited scientists. Evaluation agencies like ANVUR should probably exclude self-citations in future evaluations, for the noise introduced by the young researchers. The overall state of health of the Italian research system and the positive effect of the ANVUR assessments are demonstrated by the number of Italian researchers in the top one hundred thousand, which has increased by comparing the "career" databased of 22 years, with that of the "young" researchers in the "2017" database. Italy, looking at the elite researchers, not only is not the most indulgent in self-citations, but has shown the best improvements, proving that the introduction of ANVUR had a positive effect. Indeed, all countries apart from Italy have suffered a decline, even substantial (–20% on a national Japan scale), of the number of researchers present in the 2017 data sets compared to career data. Italy instead shows a +0.2% on a global basis and an impressive +11.53% on a national basis.

# Introduction

The recent article by Baccini et al. [1] has received large coverage in the general Italian and foreign press (see Stella in the Corriere Della Sera – https://www.corriere.it/cronache/19_settembre_11/i-professori-si-citano-soli-cosi-si-gonfia-ricerca-c471954a-d4cf-11e9-8dcf-5bb1c565a76e.shtml – in Italian), after suggesting a "boom" of self-citations of Italian scientists and therefore a deleterious effect of the evaluations of the Italian National Agency for Evaluation of the University System and Research (ANVUR) (https://www.roars.it/online/citarsi-addosso-ascesa-scientifica-

dellitalia-no-solo-doping-per-inseguire-i-criteri-anvur/ – in Italian) which has been set in Italy in 2009, for evaluating individuals, departments and universities with bibliometric parameters. This largely critical article had a greater effect on the press than positive results in the news, such as the one that indicates that in 2012 Italian publications have surpassed those in the United States in terms of weighted citation impact, or that Italy has reached second place after the United Kingdom among the G8 countries, despite the level of public spending on research and development is still (1.3%) well below the European average (1.9%) [2-4]. We therefore tried to read data starting from a different perspective of elite scientists, stemming from an even more recent article, the ranking of Ioannidis et al. [5] on PLoS Biology, which recognizes that citation metrics are widely but sometimes improperly used. Ioannidis and co-authors have published data on over 100,000 leading scientists (the term scientists here is used improperly to refer to all the components of the lists made by Ioannidis et al.) who provide standardized information on citations, eliminating self-citations and even "citation farms", thus overcoming all the problems Baccini et al. raised, although limited to elite scientists, and not the full 7.5 million scientists in the Scopus database. As Prof. Ioannidis says in a recent Nature coverage of Baccini's work, Baccini's effect is not extreme, and "could be due to chance. If the increase were real, it could be attributable to a minority of researchers" (Richard Van Noorden, Italy's rise in research impact pinned on 'citation doping', Citation of Italian-authored papers by Italian researchers rose after the introduction of metrics-based thresholds for promotions. Nature NEWS 13 SEPTEMBER 2019). This statement motivated the present study and the question we asked ourselves is: "Is it a minority or the tail of less-rated researchers (either because they are poorly performing or because they are young newcomers) trying to beat the system to rig the bibliometric evaluations through the few self-citations that they can manufacture?". Baccini et al. observe that the "inwardness" (a parameter defined by them as an index of self-reference) has undergone an increase in slope since 2009, ten years ago. Is this true also for elite scientists in Ioannidis et al database? And is this true for other countries?

This note is therefore structured as follows:

(i) the concepts of citation ratio and ranking ratio that are used in this work are defined,

(ii) therefore, an analysis is made of the citation ratio distribution and the ranking ratio of the data within the single selected sample countries;

(iii) the same analysis is then extended to the single set of countries under exam, in order to obtain the real trend of the position fluctuation in the ranking relative to the weight of the self-citations.

(iv) the study is specialized to the "Italy in the world" case, making it possible to draw the final conclusions.

Ioannidis' data are the result of a very complete analysis of the data present on Scopus, even more than the data used by Baccini on SciVal.com, as Ioannidis' career database dates back to the 1960s, providing a long-term performance of the single researchers, in order to evaluate their position in the Italian or world rankings, and the change of this position with and without self-citations. The data are separated in two databases: one for for "career" spanning a 22 years, where there are scientists of greater standing and already well-established career, and the "2017" classification, among which it is interesting to highlight both the "young promises" and the "elderly" who continue giving their contribution to the Literature.

## Data analysis for the year 2017 Ioannidis database

As already mentioned, the data used are part of a large database collected by Ioannidis et al. [2] and published in the authoritative journal PLoS Biology. In this study, the data for the single year (2017) were considered, whose file contains the data of the 106,368 most cited scientists in the world. We have considered six sample nations, or a base of 64,368 scientists, equivalent to about 60.5% of the total. The six countries examined are the United States (USA – 42,455 scientists), Great Britain (GBR – 9,467 scientists), Germany (DEU – 5,225 scientists), Japan (JPN – 2,674 scientists), Italy (ITA – 2,303 scientists) and France (FRA – 2,193 scientists). Of the large database provided by Ioannidis and co-authors, the number of citations per scientist with and without self-citations was considered in this note. The citation ratio $\rho_c$ is therefore simply defined as the ratio between self-citations and citations by others, while the ranking ratio $\rho_r$ is defined as the ratio of the difference between position in the ranking with and without self-citations and the position including self-citations, viz.:

$$\begin{aligned} \rho_c &= \text{selfcites}/(\text{cites} - \text{selfcites}) &\text{(a)} \\ \rho_r &= [\text{pos}(\text{cites}) - \text{pos}(\text{cites} - \text{selfcites})]/\text{pos}(\text{cites}) &\text{(b)} \end{aligned} \quad (1)$$

Therefore, a scientist particularly accustomed to self-referencing could have a $\rho_c$ even greater than one, while the ranking ratio is a number that, if negative, indicates a deterioration in the ranking after excluding self-citations (vice versa if it is positive).

## Analysis on a national basis

### Citation ratio

In the analysis for individual countries, we have considered the excursion in the national classification including or excluding self-citations, as well as analyzing the distribution of the citation ratio for each nation. Collecting data for each of the nations in the form of a histogram, we observed that the national distributions of $\rho_c$ remain similar (cfr. Fig 1).

**Fig 1. Trend of the citation ratio at national level for the 2017 database.** The initials WRD identify the "world" as a union of the nations in question. Axes truncated for $\rho_c \geq 1$.

From Fig 1 we note that $\rho_c$ reaches the peak value typically between 0.1 and 0.2, or between 0 and 0.1 for GBR and USA. That is, most probably an ITA, DEU, JPN and FRA scientist tends to cite himself once every seven citations received, while a GBR or US scientist will tend to be more likely to be self-cited once every around twenty citations received. Based on the trend of the data, we assumed that a two-parameter Weibull distribution could fit the trend of the citation ratios. Therefore, we calculated the parameters of the Weibull curves by applying the maximum likelihood estimate (MLE) from which we could trace Fig 2, which clearly shows that, apart from USA and GBR, ITA data are perfectly in agreement with FRA, JPN, and DEU. The global data (WRD) is obviously heavily influenced by the USA and GBR scientists, which alone make up about 80% of the statistical sample. Surprisingly, the DEU (Germany) data appear to show the highest mode, albeit being very similar compared to the others.

**Fig 2. Distribution of the country-by-country citation ratio using Weibull probability density functions.**

It is noteworthy that the data in question do not, however, consider the "very smart" (in the negative sense), i.e. those who have a citation ratio greater than one, indicating that they cite themselves more than others cite them. This case has been singularly studied and the result would seem to show a greater tendency in the ITA block to self-citation within the investigated sample, as shown in Table 1, where both the absolute number and the percentage of the analyzed national sample is shown. Therefore, in this type of analysis, a certain "Baccini effect" would seem to be found, however the conditional is compulsory since we do not know with certainty whether this data is statistically significant, being only a mere 2% of the total. In addition to this, Table 1 also shows the average position in absolute and relative terms of the "smart" scientists, highlighting the theory we put forward: typically, those who tend to self-cite are in a low position at the national level. This position fluctuates, in a normalized ranking on a percentage basis, between 73 and 92 cents with the Italians ranking around halfway with 84%.

**Table 1. Number of scientists with a citation ratio greater than one and a percentage of the national total.**

| Nation | ITA (2303) | DEU (5225) | JPN (2674) | FRA (2193) | GBR (9467) | USA (42455) |
|---|---|---|---|---|---|---|
| $\rho_c \geq 1$ | 48 | 26 | 15 | 8 | 26 | 90 |
| $\rho_c \geq 1$/ Tot | 2.08% | 0.50% | 0.56% | 0.36% | 0.27% | 0.21% |
| average[pos($\rho_c \geq 1$)] | 1944 | 4093 | 2460 | 2011 | 6941 | 34255 |
| average[pos($\rho_c \geq 1$)]/Tot | 84% | 78% | 92% | 92% | 73% | 81% |

## Ranking ratio

As for the analysis of the ranking ratio $\rho_r$, we have produced a graph for each of the six countries studied, highlighting the samples with $\rho_r<0$, i.e. in gray. those who without self-citations go down into the global ranking, and into black the samples with $\rho_r>0$. The results are shown in Fig 3. The ITA data in gray do not show significant differences to the total, in fact, all the figures tend to show a belly at $\rho_r>0$, and to widen quite evenly if not for the fact that they appear slightly more scattered.

**Fig 3. Analysis of the trend of the ranking on a national basis**. Note how the data with $\rho_r<0$ (gray) corresponding to a drop in the ranking are more scattered than the data with $\rho_r>0$ (black) identifying a rise in the ranking.

We, therefore, decided to quantify this scatter in $\rho_r$ by studying its distribution both for each nation and for the entire WRD sample. The distributions with the numeric data are shown in Fig 4, from which we note that the ratio between the number of samples in $-0.20 \leq \rho_r \leq 0$ and in $-0.40 \leq \rho_r \leq -0.2$ for ITA and DEU (Germany) is the highest of all the nations (and not Italy!). This indicates a higher worsening on average once the self-citations have been removed.

**Fig 4. Trend of the ranking ratio on a national basis**. The code WRD identifies the "world" intended as union of the nations under exam. All the distributions are centered around $0 \leq \rho_r \leq 0.2$ and show a comparable shape, confirming what observed in Fig 3.

The qualitative indication deriving from the histograms in Fig 4 is shown quantitatively in Table 2. What appears is once again a certain "Baccini effect" in the "average of worsening" line which shows how, tending to exclude self-citations, the position of people who worsen in the ranking drops on average by 20%, compared to around 10% of other nations. This effect is attenuated by weighing the figure for the ratio between the number of samples that fell in the standings compared to the total, falling to 7.5% compared to around 4.5% of foreign nations. However, it should be highlighted that the number of Italians improving on the national ranking excluding self-citations is significantly higher than those worsening, showing the highest ratio of the entire statistical base (1.65 vs. ~1.55 on average). To understand if this data is significant, we compared it with the corresponding result for Italy based on WRD.

**Table 2. Analysis of ρ_r on a national basis**

| Analysis of exclusion of self-citations | ITA (2303) | DEU (5225) | JPN (2674) | FRA (2193) | GBR (9467) | USA (42455) |
|---|---:|---:|---:|---:|---:|---:|
| risen in ranking | 1459 | 3108 | 1598 | 1297 | 5826 | 26080 |
| dropped in ranking | 884 | 2100 | 1056 | 869 | 3610 | 16351 |
| ratio(risen/dropped) | 1.65 | 1.48 | 1.51 | 1.49 | 1.61 | 1.6 |
| neutral | 10 | 0 | 20 | 27 | 31 | 24 |
| average(risen) | 7.80% | 0.00% | 6.40% | 6.10% | 5.40% | 5.10% |
| average(dropped) | -19.50% | 0.00% | -11.70% | -11.10% | -10.40% | -9.70% |
| weighted average(risen) | 5.00% | 0.00% | 3.80% | 3.60% | 3.30% | 3.10% |
| weighted average(dropped) | -7.50% | 0.00% | -4.60% | -4.40% | -4.00% | -3.70% |

## Specialization to ITA data on the WRD basis

To repeat and generalize the analysis on a "global" basis, we followed a procedure like that adopted on the national basis by combining the scientists of the individual nations into the WRD statistical base. The derived dataset of global rankings for Italian scientists has been called ITW. The results obtained do not differ significantly from the results on the ITA basis, and this was predictable, given the similarity between the various nations and the distribution of citation and ranking ratios, as shown in Figs 1-4. In Fig 5 we report the trend line showing the correlation between the ranking of researchers with $\rho_r<0$ with and without self-citing, highlighting a very slight, almost imperceptible by eye, difference between the various countries. Italy indeed shows the worst datum, both in terms of slope and of data dispersion (lowest $R^2$), but we are talking about a difference between 1.079 and 1.072, or a mere 0.55%. Finally, we observe in Fig 6 that the data on a national scale are also representative of data on a global scale since the trend of the normalized position in terms of percentages with varying $\rho_c$ for the ITA base and the ITW base are practically overlapping.

**Fig 5. Analysis of the ranking on the global basis for the only scientists having $\rho_r<0$, corresponding to a drop in the rankings.** All the regression lines are close to each other, with ITA showing a slightly more pronounced average worsening.

**Fig 6. Comparison between the normalized average (on a percent scale) ranking ITA and ITW with reducing the citation ratio threshold.** The two curves are always within the 3% scatter error bands.

# Discussion

## The career data

To substantiate the findings and to confirm that there is not a trend in time of increasing the self-citations amongst the best performing scientists of Italy, a comparative analysis with the "career" database from Ioannidis et al. is required. If the $\rho_c$ distribution tends to higher values in time, then the "Baccini effect" is demonstrated, otherwise, it is confuted. The same procedure described for the "2017" data has been adopted here for the career data, which presumably indicate a backward in time analysis, and the plots shown in Figs 7 - 9 have been obtained. As it is clear from Figs 7 and 8, the ITA index of self-citations $\rho_c$ does not show any alarming increase in time, since the 2017 distribution is shrunk with respect to the career and the mode is slightly decreased. Furthermore, DEU and JPN both look very similar to ITA, whilst FRA seems to show an increase in self-referentiality in time. GBR and USA data show the lowest values of $\rho_c$, but this effect is most probably related to the huge amount of citations those scientists that receive rather than the low number of self-citations, since those two nations lead by far both the rankings. Hence, from Figs 7 and 8 maybe the more alarming effect of the analysis is the increase of $\rho_c$ in FRA, which moreover combined with WRD 2017 for FRA seeing a 9% loss on the national base with respect to the WRD career database. Concerning Fig. 9, it is practically equal to Fig 4, confirming that the 2017 database does not differ significantly from the career.

**Fig 7. Comparison in the distribution of the citation ratios $\rho_c$ between career and 2017.** Markers indicate the 2017 and a solid line the career. ITA is in blue.

**Fig 8. Trend of the citation ratio at national level for the career database.** The histograms are not truncated to $\rho_c \geq 1$, since no-one of the top 100,000 scientists in its entire career has recurred to self-citations more times than he has been cited.

**Fig 9. Trend of the ranking ratio on a national basis for the career database.** All the distributions are centered around $0 \leq \rho_r \leq 0.2$ as in the 2017 database.

In light of the demonstrated fact that the effect of self-citations on highly cited scientists is negligible and with the aim of highlighting the overall state of health of the Italian research system and the probably positive effect of the ANVUR assessments, we show in Table 3 how the ITA presence has increased in the ranking of Ioannidis et al. comparing the quantity of scientists in the 2017 database, with respect to the career database, both in terms of absolute and percent increments. The emerging facts are two:

1. all the countries apart from Italy have suffered a decline, even substantial (–20% on a national JPN scale) of the number of samples present in the 2017 data sets compared to career data. Italy instead shows a +0.2% on a global basis and an impressive +11.53% on a national basis;

2. the total number of WRD scientists in the top 100,000 suffered a significative reduction in the ranking, passing from 69,114 to 64,497 and with a relative reduction of 6.68%. This aspect is critical and would need further investigation, probably by including the nations that are making greater improvements in the last years, such as China.

**Table 3. Relative and absolute analysis of the presences by nation in the Career and 2017 databases**

| Presence in the rankings | | ITA | DEU | JPN | FRA | GBR | USA | WRD |
|---|---|---|---|---|---|---|---|---|
| Absolute presence | Career data | 2,065 | 5,459 | 3,382 | 2,415 | 9,780 | 46,013 | 69,114 |
| | 2017 data | 2,303 | 5,225 | 2,674 | 2,193 | 9,647 | 42,455 | 64,497 |
| Percent presence | Career data | 1.97% | 5.20% | 3.22% | 2.30% | 9.31% | 43.81% | 65.81% |
| | 2017 data | 2.17% | 4.91% | 2.51% | 2.06% | 9.07% | 39.91% | 60.64% |
| | National Δ | 11.53% | -4.29% | -20.93% | -9.19% | -1.36% | -7.73% | -6.68% |
| | Global Δ | 0.20% | -0.29% | -0.71% | -0.24% | -0.24% | -3.90% | -5.17% |

The union of the nations, WRD, has been considered as a nation itself in order to analyze also its trend.

## Conclusions

The effect claimed by Baccini et al. that due to ANVUR's bibliometric evaluations of individuals, departments, and universities, stating that in Italy there has been a surge in self-citations in the last ten years, since extremely limited, has been denied regarding the data of the hundred thousand most "highly cited" scientists in the world, including about two thousand Italians. We have found that only those who are in a low position at the national level have a slight tendency to self-cite a lot. While we obviously advise evaluation agencies such as ANVUR to exclude self-citation in future assessments, particularly of young researchers or low rank researchers (which is very simple to do in all databases), we provide a set of original conclusions. Indeed, by comparing the "career" database of Ioannidis et al, with that of the young scientists "2017" database, we have found a much more important fact: all the countries we studied apart from Italy have suffered a decline, even substantial (–20% on a national JPN scale) of the number of scientists. Italy instead shows a +0.2% on a global basis and an impressive +11.53% on a national basis.

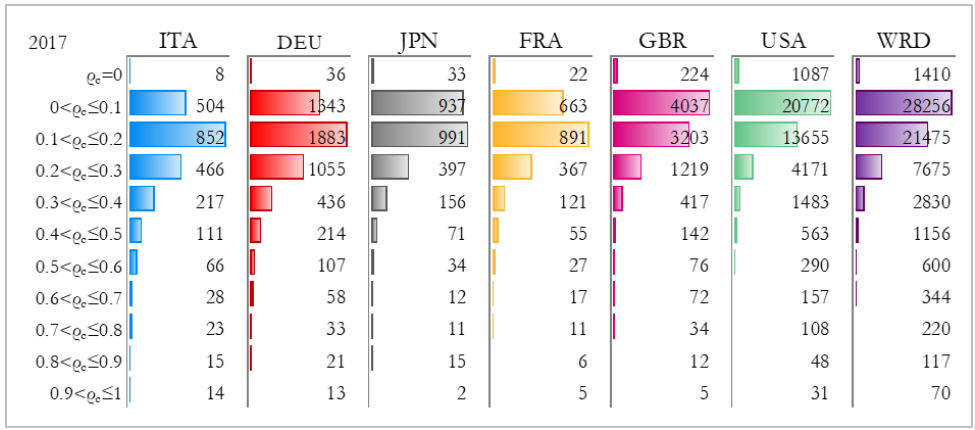

**Fig 1. Trend of the citation ratio at national level for the 2017 database.** The initials WRD identify the "world" as a union of the nations in question. Axes truncated for ρ_c≥1.

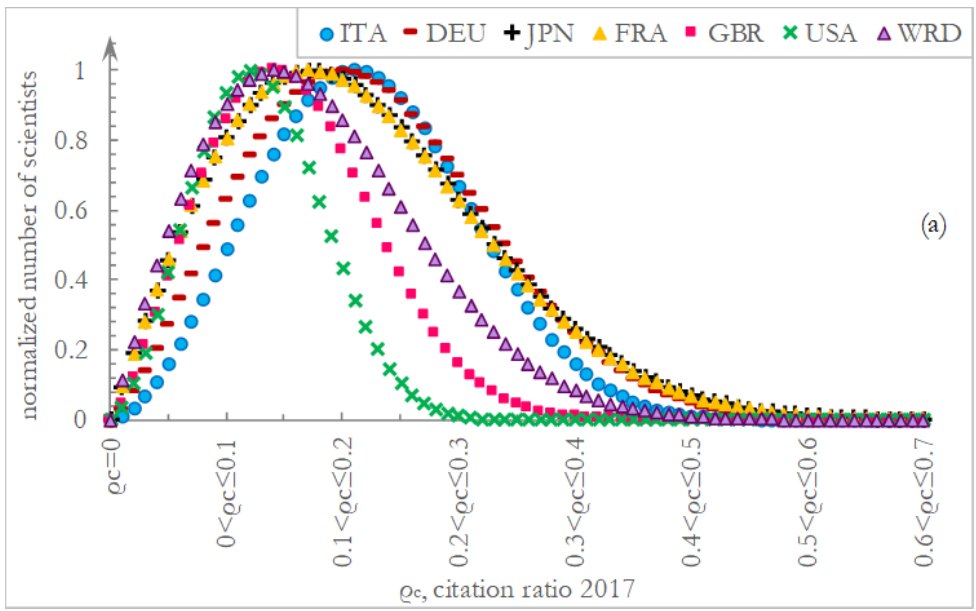

**Fig 2. Distribution of the country-by-country citation ratio using Weibull probability density functions.**

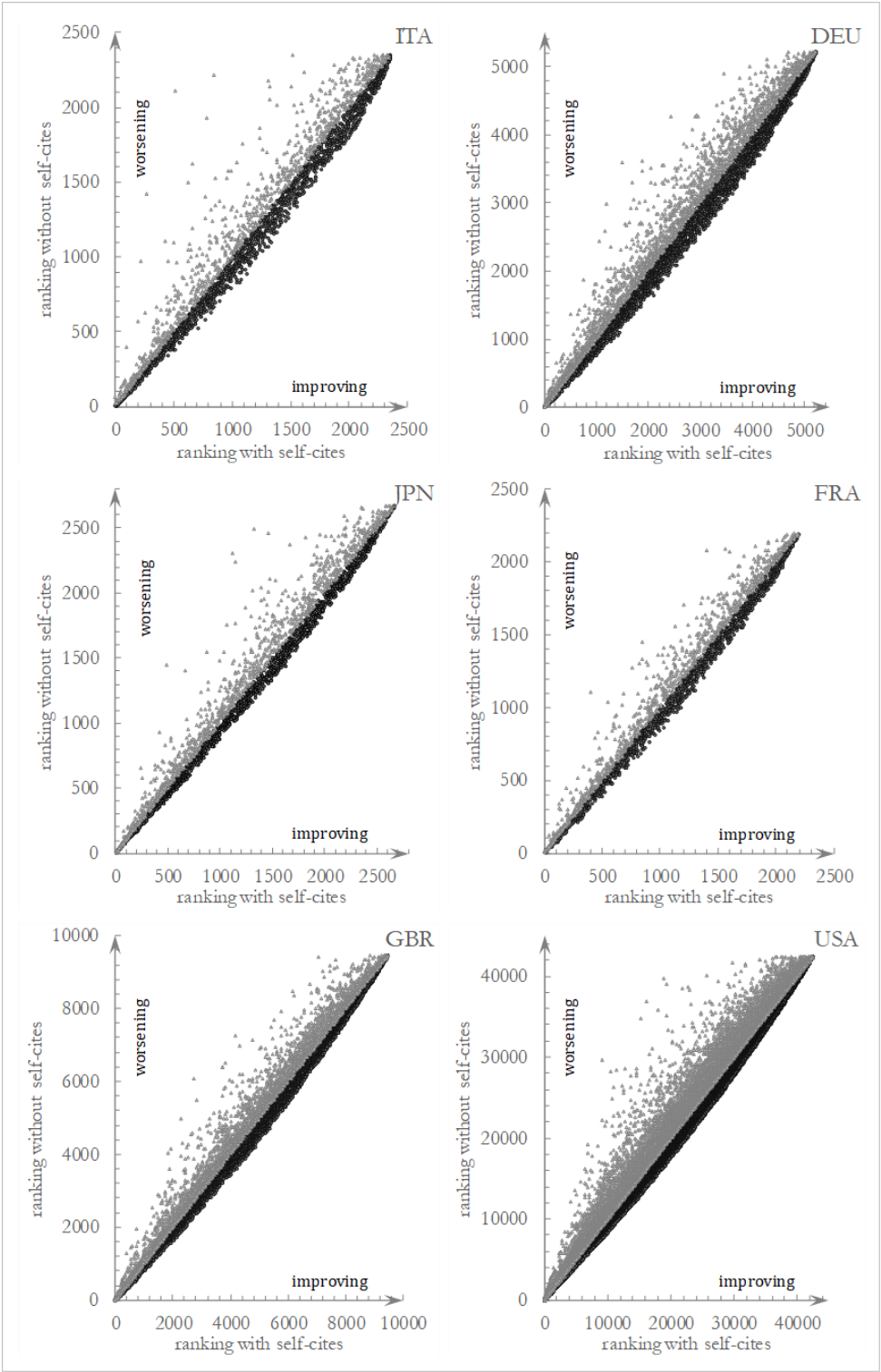

**Fig 3. Analysis of the trend of the ranking on a national basis.** Note how the data with $\rho_r<0$ (gray) corresponding to a drop in the ranking are more scattered than the data with $\rho_r>0$ (black) identifying a rise in the ranking.

| 2017 | ITA | DEU | JPN | FRA | GBR | USA | WRD |
|---|---|---|---|---|---|---|---|
| $\rho_r \leq -1$ | 27 | 30 | 7 | 7 | 28 | 120 | 233 |
| $-1 < \rho_r \leq 0.8$ | 7 | 17 | 2 | 3 | 22 | 45 | 97 |
| $-0.8 < \rho_r \leq -0.6$ | 17 | 25 | 17 | 10 | 38 | 105 | 202 |
| $-0.6 < \rho_r \leq -0.4$ | 37 | 68 | 30 | 23 | 66 | 360 | 636 |
| $-0.4 < \rho_r \leq -0.2$ | 95 | 238 | 109 | 66 | 312 | 1259 | 2149 |
| $-0.2 < \rho_r \leq 0$ | 711 | 1739 | 911 | 787 | 3175 | 14486 | 21505 |
| $0 < \rho_r \leq 0.2$ | 1395 | 2980 | 1578 | 1269 | 5755 | 25985 | 39254 |
| $0.2 < \rho_r \leq 0.4$ | 59 | 128 | 20 | 28 | 71 | 95 | 291 |
| $0.4 < \rho_r \leq 0.6$ | 5 | 0 | 0 | 0 | 0 | 0 | 0 |
| $0.6 < \rho_r \leq 0.8$ | 0 | 0 | 0 | 0 | 0 | 0 | 0 |
| $0.8 < \rho_r \leq 1$ | 0 | 0 | 0 | 0 | 0 | 0 | 0 |

**Fig 4. Trend of the ranking ratio on a national basis**. The code WRD identifies the "world" intended as union of the nations under exam. All the distributions are centered around $0 \leq \rho_r \leq 0.2$ and show a comparable shape, confirming what observed in Fig 3.

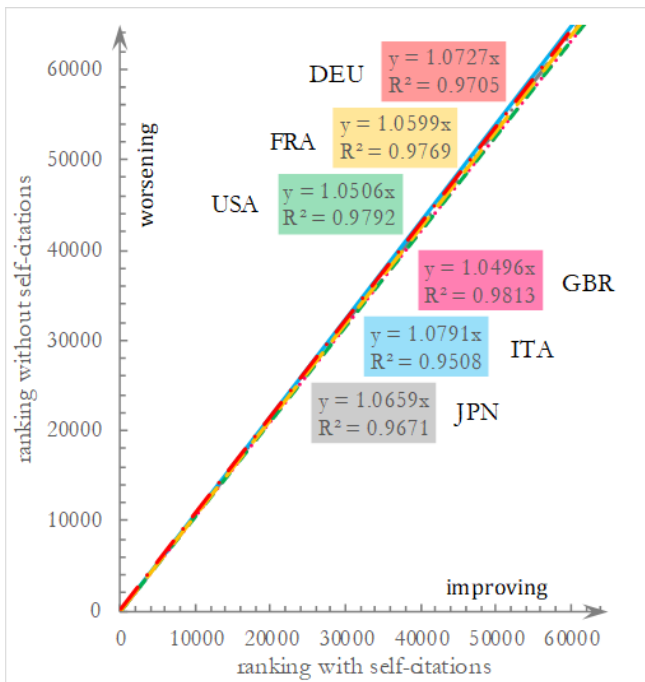

**Fig 5. Analysis of the ranking on the global basis for the only scientists having $\rho_r < 0$, corresponding to a drop in the rankings.** All the regression lines are close to each other, with ITA showing a slightly more pronounced average worsening.

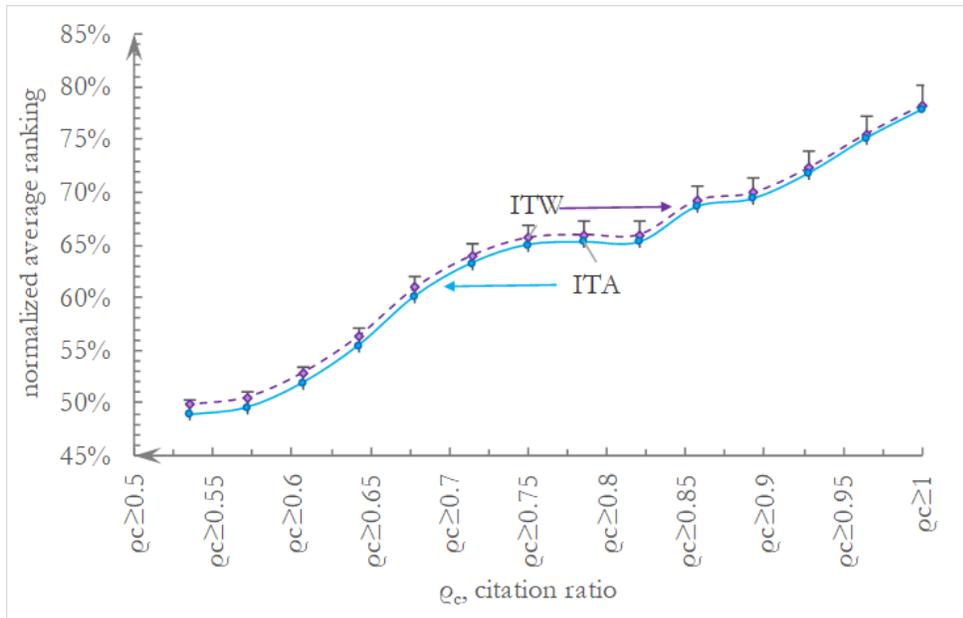

**Fig 6. Comparison between the normalized average (on a percent scale) ranking ITA and ITW with reducing the citation ratio threshold.** The two curves are always within the 3% scatter error bands.

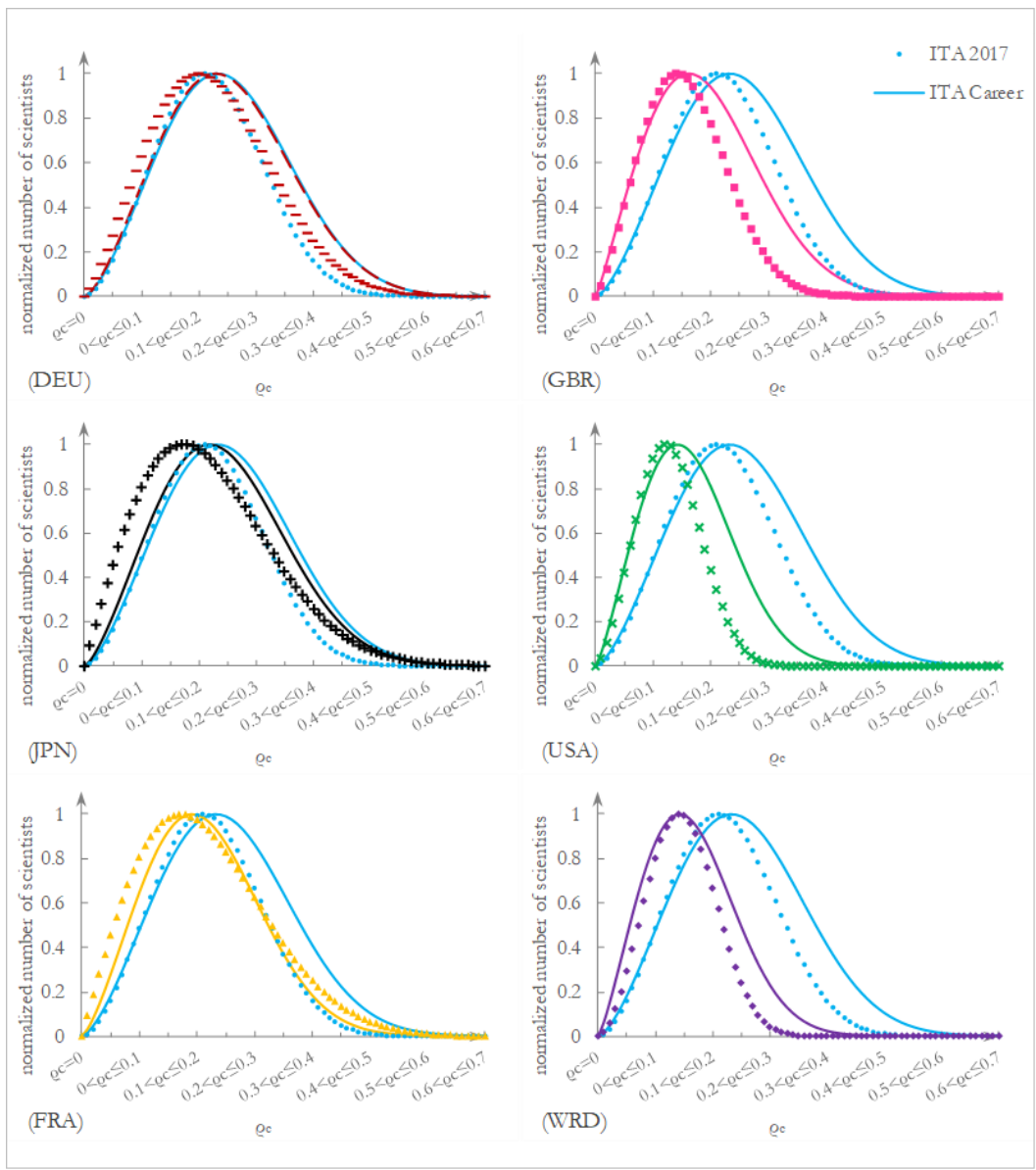

**Fig 7. Comparison in the distribution of the citation ratios $\rho_c$ between career and 2017.**
Markers indicate the 2017 and a solid line the career. ITA is in blue.

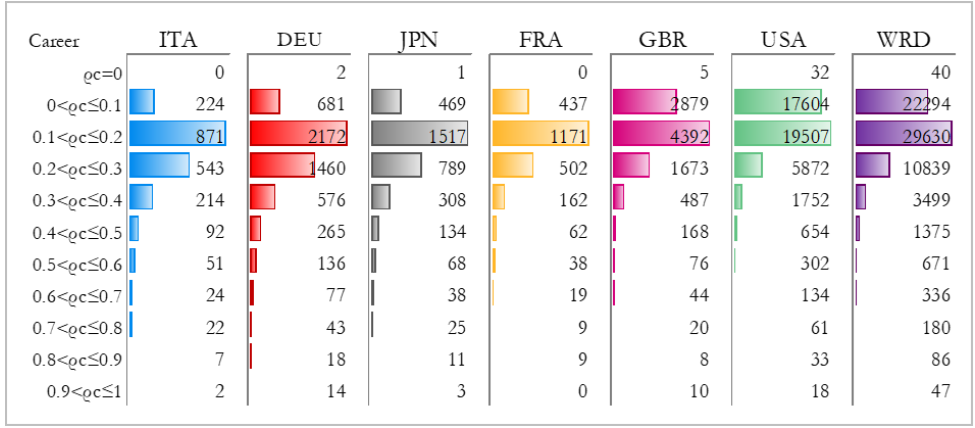

**Fig 8. Trend of the citation ratio at national level for the career database.** The histograms are not truncated to $\rho_c \geq 1$, since no-one of the top 100,000 scientists in its entire career has recurred to self-citations more times than he has been cited.

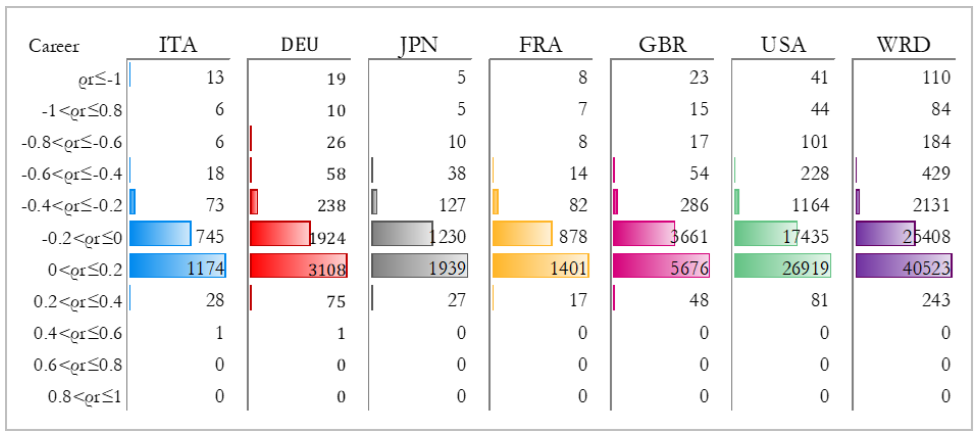

**Fig 9. Trend of the ranking ratio on a national basis for the career database.** All the distributions are centered around $0 \leq \rho_r \leq 0.2$ as in the 2017 database.